\documentclass[12pt,preprint]{aastex}

\shorttitle{Structures in WR Winds}
\shortauthors{Marchenko et al.}

\begin{document}

\title{Origin of Structures in Wolf-Rayet Winds: {\it FUSE} Observations of WR 135\altaffilmark{1}}

\author{S.V. Marchenko\altaffilmark{2}, A.F.J. Moffat\altaffilmark{3},
 N. St-Louis\altaffilmark{3}, A.W. Fullerton\altaffilmark{4,5}}

\altaffiltext{1}{Based on observations made with the NASA-CNES-CSA {\it
{Far Ultraviolet Spectroscopic Explorer. FUSE}} is operated for NASA by
The Johns Hopkins University under NASA contract NAS5-32985.}
\altaffiltext{2}{Department of Physics and Astronomy, Western Kentucky
University, 1906 College Heights Blvd., 11077, Bowling Green, KY
42101-1077, USA; sergey.marchenko@wku.edu}
\altaffiltext{3}{D\'epartement de Physique and
Observatoire du Mont M\'egantic, Universit\'e de Montr\'eal, CP 6128,
Succursale Centre-Ville, Montr\'eal, QC H3C 3J7, moffat,stlouis@astro.umontreal.ca }
\altaffiltext{4}{Department of Physics and Astronomy, University of
Victoria, P.O. Box 3055, Victoria, BC, Canada.}
\altaffiltext{5}{Center
for Astrophysical Sciences, Department of Physics and Astronomy, The
Johns Hopkins University, 3400 North Charles Street, Baltimore, MD
21286, USA; awf@pha.jhu.edu}

\begin{abstract}
We report the detection with FUSE of strong, highly blue-shifted
absorption features appearing in the absorption troughs of practically
all major P Cygni profiles in the presumably single Wolf-Rayet star
WR135. These features also appear in the shock-sensitive OVI 1032/38
\AA\ doublet, coincident both in time and velocity space with the rest
of the lower-ionization species. Choosing between two alternative
interpretations: large-scale, coherent structures vs. localized, random shocks,  we
favor the latter. The absolute value of
velocity, as well as velocity dispersion in the
shocked region, the density of the shocked gas, and the time
scales of the observed variability allow us to relate the observed shocks to the
incidence of numerous over-dense clumps (blobs) in the wind of a hot, massive star.

\end{abstract}


\section{Introduction}

All massive ($M_{init} \sim$20-100 $M_{\odot}$) stars drive
strong winds.  These stars spend most of their lifetimes in the hot part
of the HRD, either as O (H-burning) or Wolf-Rayet (WR: mainly
He-burning) stars.  Although hot-star winds were suspected to be
inhomogeneous (e.g. Cherepashchuk et al. 1984), truly direct evidence
came only from the detection of stochastic, systematically moving
subpeaks on the strong emission lines of WR stars (Moffat et
al. 1988). This profoundly changed the approach in atmospheric modeling of WR stars
(e.g. Hillier 1991; Hamann \& Koesterke 1998, Hillier \& Miller 1999)
and resulted in a substantial reduction, by a factor 2-5, of predicted mass-loss rates.
A similar factor may also apply to O-star winds (e.g. Eversberg et al. 1998, Donati et
al. 2002, Fullerton, Massa \& Prinja 2005).
Since the evolution of massive stars
is heavily affected by mass-loss, such a significant reduction in
mass-loss rate called for a major revision in models of massive stars
and stimulated further research (see Meynet \& Maeder 2005, and
references therein).

Although the consequences of wind inhomogeneities for mass-loss estimates
are understood, the origin of the clumps is not.  It is almost certain that radiative instabilities
(e.g. Owocki 1994) must ultimately lead to supersonic,
compressible turbulence (Henriksen 1994).  Then it is reasonable to
assume that the resulting shocks will generate the X-ray fluxes seen
emanating (e.g. Berghoefer et al. 1997, Wessolowski 1996) from all hot-star
winds.
Linking shocks to X-rays,
one should look for a stochastic variablility of X-ray flux,
expecting that rather infrequent shock-shock collisions
are the most influential events in X-ray production
(Feldmeier et al. 1997). However, no X-ray facility is sensitive enough
yet to  detect such fluctuations (Bergh\"ofer et al. 1997).  In addition, the situation
is complicated by the large optical depth of the outflows from WR
stars. While the clumps are routinely observed at the base of the wind
$r ^< _\sim 100 \, R_\star$ where the emission lines arise, the
[presumably related] X-ray flux emerges from $r \gg 100 \, R_\star$.

In order to establish that embedded shocks are responsible for the
clumpy structure of hot-star winds, one may rely on observations of
resonance lines of "super-ions," i.e., ions whose presence cannot
be explained by photoionization in winds characterized by
temperatures between 20 and 40 kK (e.g. MacFarlane
et al. 1994).
The best superionization feature is the strong OVI resonance line with
$\chi_{ioniz} =$ 114-138 eV at
1031.93/37.62 \AA\ , i.e. accessible only with {\it {Far Ultraviolet
Spectroscopic Explorer (FUSE)}}.  If
the OVI line
feature is formed in hot, shocked gas filaments (clumps) permeating the
accelerating wind, one should see a variable profile, with corresponding
statistical properties (velocity dispersion, timescales,
amplitudes) resembling those found in optical emission lines,
presumably produced via superposition of counltless clumps (L\'epine \&
Moffat 1999). One may also expect, by analogy with the optical lines, the non-saturated
P Cygni absorption parts of the profiles to show the highest level of variability.


Our target, the presumably single, luminous Wolf-Rayet star WR135 = HD
192103 (WC8) is known to show a highly variable, structured (clumpy)
wind (L\'epine et al. 2000), as well as being an X-ray source (ROSAT: $2.1\pm 1.7 [10^{-3} \,
cts \, s^{-1}]$, although barely exceeding a $1\sigma$ detection limit - see
arguments in Oskinova et al. 2003). In this star, as with virtually all
other WR stars observed in the optical (L\'epine \& Moffat 1999), spectral
subpeaks are seen to propagate outwards towards the edges of strong emission
lines.
These peaks are interpreted as density enhancements emitting
excess recombination-line radiation, propagating on average along with
the general wind outflow. In WR 135 the moving subpeaks typically range
up to 0.2-2.0\% of the underlying emission-line strength, have a radial
velocity dispersion $\sigma _r = 175 \, \, km\, s^{-1}$, and last for 7.5
hours on average (L\'epine et al. 2000). The spectrum of WR135, leaving aside
the prominent O VI feature, can be modeled with $T_{eff} (\tau
_{Ross}=2/3) = 2.7 \, 10^4$ K (Dessart et al. 2000).
On the other hand, the O VI doublet, the central
subject of our discussion, peaks in strength at $T\sim 3\, 10^5$ K
(Zsarg\'o et al. 2003,
 Sembach et al. 2004). This high ionization, two stages above the dominant level
(Cassinelli \& Olson 1979), is
explained by Auger ionization in the presence of X-rays generated in
numerous wind-embedded shocks.

Here we report the simultaneous appearence of short-lived (hours)
absorption features in the high-velocity, blueshifted flanks of all
major P Cygni emission profiles (OVI included) in the {\it {FUSE}}
range. We link these absorption features to strong, highly supersonic shocks
embedded in the wind of WR 135.

\section{{\it {FUSE}} Observations}

We observed WR 135 with {\it FUSE} (Moos et al. 2000, Sahnow et
al. 2000) in October 2004, starting at HJD 2453287.5930 (UT=02:11, Oct. 9), and ending at
HJD 2453288.8483 (UT 08:19, Oct. 10), thus following the star for 30+ hours practically
non-stop.  The spectra were obtained in histogram (HIST) mode, with the
MDRS ($4"\times 20"$) aperture, covering 910-1185 \AA\ .  Initially, 59
exposures (root name D0440101\*) were scheduled back-to-back over 18
consecutive orbits. However, for technical reasons, only 42 spectra were
obtained, with a particularly large, $\sim 4$h gap during planned
exposures 15-29, when detectors autonomously shut down.
Exposures 5 and 55 were skipped because of acquisition problems.
The majority of exposures were taken with $T_{exp} \sim
500\, s$. The main setback came from a mis-alignment of the SiC
channels causing loss of flux. However, the LiF channels performed well.
The latest version of CalFUSE (v3.0.7) was used to extract and
calibrate spectra from the two LiF channels.

Armed with a large number of
spectra of the same object obtained within a short time interval,
we tried to maximize the signal-to-noise ratio (S/N) of
individual exposures.
The presence of strong,
frequently saturated IS absorptions called for minimal usage of any
binning in velocity space on the original spectra. However, it did not
preclude an appropriate smoothing of the differences (individual minus
average, individual minus individual).  We
assembled the spectra into groups, combining together the A and B
segments of the 2 LiF detectors into 4 separate averages and then
inter-comparing the averages.  Before combining the spectra, we
corrected all small-scale deviations ($< 3\, km\,s^{-1}$) in the rest-frame radial velocities
of strong interstellar absorption lines.
By comparing the averages, we found that the LiF2A and
LiF1B segments have practically the same S/N, although the LiF1B flux
was reduced by 10\%. On the other hand, LiF2B has a S/N that is two
times lower than that of LiF1A. Hence, initially we treated all 4
segments separately, then appropriately combined the processed LiF2A
with LiF1B and LiF2B with LiF1A at the very end of the procedure. This
also provided an efficient line-by-line cross-check of the suspected
variability patterns.
Experimenting with various grouping algorithms,
we finally decided to average all available
spectra with comparable S/N levels, taking the
median values and additionally rejecting at each given wavelength the 10 most
deviating flux readings (5 maximal and 5 minimal). Then the average was
subtracted from each individual spectrum, channel-by-channel, and the
difference was smoothed with a Gaussian filter. Subtracting the smoothed
difference from the former (thus removing the low-frequency instrumental trends but retaining the
high-frequency signal), and then
subtracting this second-order difference from the original spectrum (thus reducing the
high-frequency noise), we managed to
increase the S/N in each {\it individual} spectrum from the initial $\sim
15$ to $\sim 40$, without degrading the spectral resolution. Finally, by
carefully assessing the spectrum-to-spectrum variability, we found that the
processed spectra can be assembled into 3 groups (Fig. 1): exposures 1-14
(mid-time HJD 2453287.6664), 30-39 (HJD 2453288.1803)
and 40-59 (HJD 2453288.6066).

\section{Results and Discussion}

Despite the overall improvement in S/N, our detection limits do not
exceed 5-10\% (refering to the ambient continuum), depending on the spectral range,
in any
individual exposure. Fixed-pattern noise, i.e. small-scale and fairly
stable, within a few hours, noise patterns are mainly responsible for the
problem. These detection limits are $\sim$ a factor of two lower in the
grouped exposures.  Within these limits the emission parts of all major
transitions can be considered as constant. Recall that in the optical
lines the variability does not exceed $\sim 2\%$ (L\'epine et al. 2000).
In Fig. 1 we show the major variable part - that of the absorption troughs
of the P Cygni profiles. To normalize the velocities, we used
the terminal wind velocity $v_\infty=1479\, km\,s^{-1}$ from Willis et
al. (2004), which fits the data better than $v_\infty=1343\pm496\,
km\,s^{-1}$ from Niedzielski \& Skorzynski (2002).

Finally we converted the detected variations to the differences of apparent optical depths,
$d=ln(F1/F2)$ (Savage \& Sembach 1991), where F1 is an average of exposures 1-14, and F2 is
either an average of exposures 30-39 or an average of exposures 40-59 (see Fig.1).
To reduce the noise, we removed all
artefacts introduced by the saturated features (literally, the results
of 0/0 division) and smoothed the ratio F1/F2 using a gaussian filter
with FWHM=15 pix. The smoothing does not change the original FWHM values
of the features, however slightly ($_\sim ^< 10\%$) reduces their
depths. One immediately notices (Fig. 2, exposures 1-14 vs. exposures 40-59)
the high-velocity, $v\approx -1.5 \rightarrow - 1.0 v_\infty $ feature
seen in all well-developed P Cygni
profiles across the spectrum. The excellent match between these features
seen in both components of the S IV doublet lends additional credibility
to the result. This absorption component appears to be much wider in the CIII
line, which can be explained by the multiplet nature of this transition, with contributing
components being separated by as much as $\sim 0.1\, v_\infty$. While preparing Fig. 2,
we tentatively assigned $\lambda _0 = 1175.7\AA $ to C III, noticing that the components
are spread between $\lambda \lambda = 1174.93-1176.37\AA$.
Another prominent feature at $v\approx-0.8 \rightarrow - 0.2 v_\infty $
is seriously affected by saturation (OVI, S IV and CIII
profiles), which causes the displacements, or even near disappearance
of this feature. However, this absorption component is clearly seen in the non-saturated
Si III profile. From now on we focus our attention on the feature less
affected by saturation at $v\approx-1.5 \rightarrow  - 1.0 v_\infty $, which,
being  above $v_\infty$, is practically free from the ambiguity
introduced by the underlying, forward-scattering emission component.

We notice the symmetric
appearance of this blueshifted absorption in S IV, as expected for a feature originating
in an optically thin region with radially isotropic (backward/forward) turbulence.
The asymmetry, the blueshift
and the larger extension of the CIII ( and, to a lesser
extent, O VI) feature can be related to the high optical depth (i.e., substantial
self-absorption) of the transition and saturation
of the relevant part of the profile: compare S IV or O VI to
CIII  at $v ^> _\sim  -1.15 v_\infty$ in Fig. 1.

Apparently, the velocity spacing of the
OVI doublet places the variable ($v<-1.2v_\infty$) part of the absorption trough of the
1037.6\AA$\,$  component (Fig. 1) on the $v<-0.1 v_\infty$ emission part of the
1031.9\AA$\,$  line,
which may complicate the interpretation, if one assumes that the observed variability
could be caused by changes in the underlying emission. For the OVI 1037.6\AA$\,$
transition the detection of variability at  $v<-0.2v_\infty$ is impaired by the
presence of strong, saturated IS absorptions.  However, we remedy with the
CIII 1175.7\AA$\,$  complex. Its strong, well-defined variability at $v<-1.2v_\infty$
(Fig. 1) is not accompanied by any detectable changes in the    $-0.4v_\infty<v<-0.1v_\infty$
part of the profile (Fig.2).

There are two possible explanations for the detected blue-edge ($v>v_\infty$) variability:
discrete absorption components (DACs) and fairly localized, wind-embedded
shocks. The DACs are thought to be related to the co-rotating interaction
regions (CIRs: Cranmer \& Owocki 1996), the zones where high-speed
streams generated by disturbances at the base of the wind collide with relatively slower
parts of the wind.  One should notice the high incidence of DACs in the winds of OB stars
(Kaper et al. 1996) vs.  relative rarity of
these large-scale, spatially-coherent structures in the WR winds (see
Marchenko 2003 and references therein).

Have we captured a single, isolated, stochastically appearing shock or a CIR-like
structure which periodically crosses the line of sight? We favor the former, as there is no
trace of any large-scale, coherent structures in the optical (L\'epine
et al. 2000). Plus, the wind of WR135 is not flattened, contrary to all the
CIR-bearing examples: WR6, WR134 and WR137 (see Lef\`evre et al. 2005 and
references therein). In OB stars any CIR-related variability takes place in
a wide velocity range, $-v_\infty <v < v_\infty$, practically never
exceeding $v_\infty$ (rather, slightly modulating the blueshifted absorption edge:
Kaper \& Henrichs 1994).
In WR135 we see a clear dominance of
variability which goes far beyond $-v_\infty$, by as much as 50\%.
And, we see no signs of variability exceeding the 5-10\% detectability levels
in the `traditional' CIR domain, $-v_\infty <v < v_\infty$ (cf. the well-studied
case of WR6: St-Louis et al. 1995). In WR135, rather than modulating the $-v_\infty <v$
edge, as usually attributable to CIRs, the short-living additional absorption is perceived
as an isolated feature (cf. S IV 1073 in Fig. 1) with no signs of a gradual shift to higher velocities
(cf. CIII 1175.7  in Fig. 1).
In addition, the timescales of the detected UV variability, $\sim 0.5-1.0$ d,
correspond reasonably well to the average `survival' times of blobs in
the optical region, $ _\sim ^< 0.5$ d (L\'epine et al. 2000).  And, the FWHMs
of the features seen in Fig. 2, $FWHM\approx 400 \,
km\,s^{-1} \approx 2.4 \sigma$ (assuming a Gaussian profile) closely
match the observed value of the radial velocity dispersion in the blobs
seen in optical emission lines, $\sigma = 175 \, km\,s^{-1}$.
The synchronized
appearence of the absorption features in all major P Cygni profiles
across the {\it FUSE} spectrum mimics the general correlation in
the behavior of blobs observed in different lines in the optical (L\'epine
et al. 2000, Lef\`evre et al. 2005).
Hence, one may relate the
observed traces of shocked (and rapidly cooling, thus getting denser)
gas to the incidence of numerous over-dense clumps (blobs) in the wind
of WR135. Of course, the great spatial coherence of a DAC-related structure may
facilitate detection. On the other hand, while adopting the model of a structured
(clumpy) WR wind, one must count on the presence of a large amount
of profile-forming, overdense structures (e.g. Lepine et al. 2000).
The sheer number of stochastically distributed inhomogeneities
leads to high probability of seeing, at any
given moment, such a structure in absorption (i.e., projected on the optically thick
part of the wind).

One may provide additional arguments favoring the `shock' scenario over
DAC (CIR)-related variability.
There are numerous examples of $v>v_\infty$ variability in WR stars
observed by {\it IUE}: WR 136 (St-Louis et al. 1989), WR 6 (St-Louis et
al. 1995), to name a few. An extensive search for OVI variability in the
winds of OB stars (Lehner et al. 2003) shows that a solid majority,
at least $64\%$ of the O3-B1 targets,
can be considered as variable.
However, {\it none} of the surveyed objects showed $v>v_\infty$
variations in  O VI, despite the tendency of {\it {stronger shocks to occur more
frequently near $v_\infty$}}, and despite the tendency of more than 80\% of O stars
(Howarth \& Prinja 1989) to show DACs in their spectra. Hence, the O VI line seems to
be less sensitive to the DAC-related phenomena in comparison to other strong UV transitions.
The only notable exception is $\alpha\ Cam$,
where the detected variability in UV resonance lines (Lamers et
al. 1988) extends far beyond the derived $v_\infty = 1550\, km\,s^{-1}$
(Repolust et al. 2004).

If we are to interpret the variability as some kind of relaxation of the
initially `unperturbed' profile (red line in Fig. 1), then the OVI
profile has the highest restoration pace.
This could be related to fast cooling of the shocked gas. Also, note
that in the CIII and SIV profiles the
higher-velocity (thus presumably higher-temperature) region is restored much
faster.
Adopting relative shock speeds of $\Delta v = v_{shock} - v_\infty$ $^<
_\sim 10^3 km\,s^{-1}$, one obtains $T\sim 3 \, 10^7$ K for the shocked
gas (e.g. Stevens et al. 1992). Assuming that the cooling time is
related to the disappearence of the extra-absorption features at $v >
v_\infty$ (Fig. 1), $t_{cool}\sim 10^5$ s, and taking the cooling
function for the WR wind (e.g. St-Louis et al. 2005), we estimate the
density of the shocked gas as $n\sim 10^9 cm^{-3}$. Hence, we observe shocks
coming from the same region populated by blobs: $r = 6...82\,R_\star$ corresponds to
 $n=10^{10}...10^8\, cm^{-3}$ in the wind of WR135 (A. Barniske, priv. comm.).
Indeed, taking the flow
time $t=R_\star/v_\infty = 20 \, min$ (cf.  Koesterke et al. 2001) and
assuming that blobs originate at the base of the wind and survive for
$^< _\sim 10\, h$ (on average, for 7.5h: L\'epine et al. 2000), we find
that the blobs observed in the optical populate the region out to  $r _\sim ^<
30\,R_\star$, in fair agreement with the estimate coming from
the UV.

\section {Conclusions}

From our simultaneous detection ({\it both in time and velocity space})
of a short-lived (hours) absorption feature in the $v>v_\infty$ parts of
the emission profiles of lines with a wide range of ionization
potentials, including the shock-sensitive OVI doublet, we can conclude
that:

(a) in the winds of hot, massive stars the highly blueshifted, $|v| >v_\infty$,  parts of
the P Cygni profiles may originate in high-speed, shocked regions of the winds;

(b) considering the enormous, $\sim 1/4 v_\infty$, FWHM of the features (Fig. 2), one can safely
assume that a network of shocks, each with an adequate azimuthal extension,
may create a flow with highly non-monotonic velocity field, thus resulting
in a completely saturated P Cygni trough - a hypothesis voiced a long
time ago by Lucy (1983; with further developments in Puls et al. 1993), but never proved
with any certainty;

(c) the observed shocks may be responsible for the relatively soft X-ray
flux from OB/WR winds, with relative (to an unperturbed flow) velocities of the shocked gas
reaching $^< _\sim 500\, km\,s^{-1}$, thus providing $kT^< _\sim 0.5
keV$, in good agreement with the characteristic X-ray temperatures of
single OB stars (Bergh\"ofer et al. 1996);

(d) if shocks and blobs are related, then,
considering the propagation of reverse/forward shocks (now the same as
blobs) in the wind, we cannot expect them to follow the velocity law of
an unperturbed wind. This may explain the seemingly low acceleration
of blobs in the WR winds (see the discussion in Koesterke et al. 2001).

\acknowledgments

We thank Stan Owocki for enlightening discussions of the wide range of
issues related to wind instabilities. This work was partially supported
by the NASA research grant NAG5-13716.  AFJM and NSL are grateful for financial
assistance from NSERC (Canada) and FQRNT (Quebec). We
thank the first referee for his constructive criticism which helped to streamline
our discussion.

{\it Facilities:} \facility{FUSE}.

\clearpage

\begin{figure}
\epsscale{1.0}
\plotone{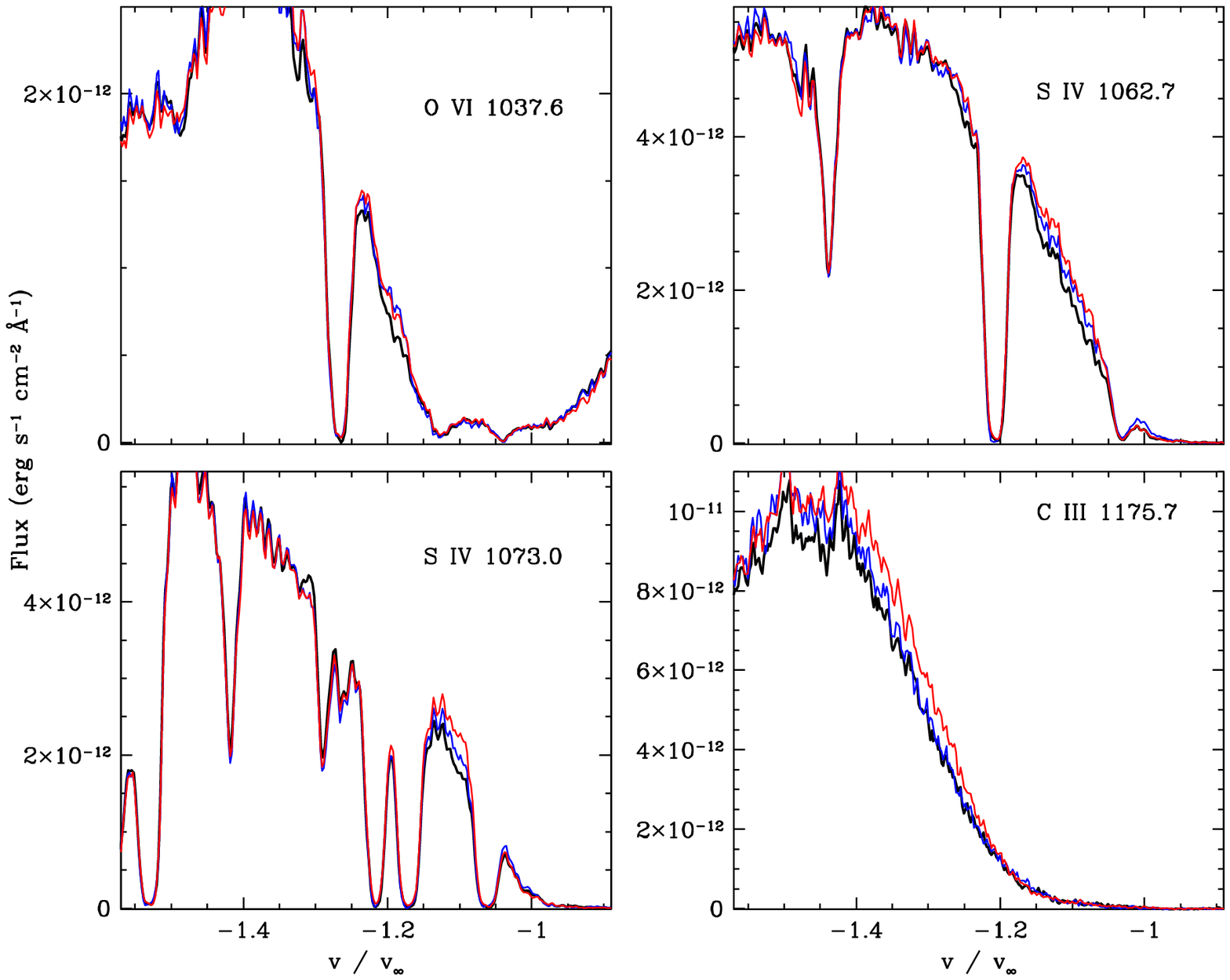}
\caption{
Fig. 1. The average profiles of major emission lines
grouped as: UT 2:11-5:35 (exp. 1-14, black lines), UT 13:31-15:35 (exp. 30-39, blue
lines)
and  UT 20:34 - 08:11 (next day, exp. 40-59, red lines), i.e. with time progressing from
black to red.\label{fig1}}
\end{figure}

\clearpage

\begin{figure}
\epsscale{.60}
\plotone{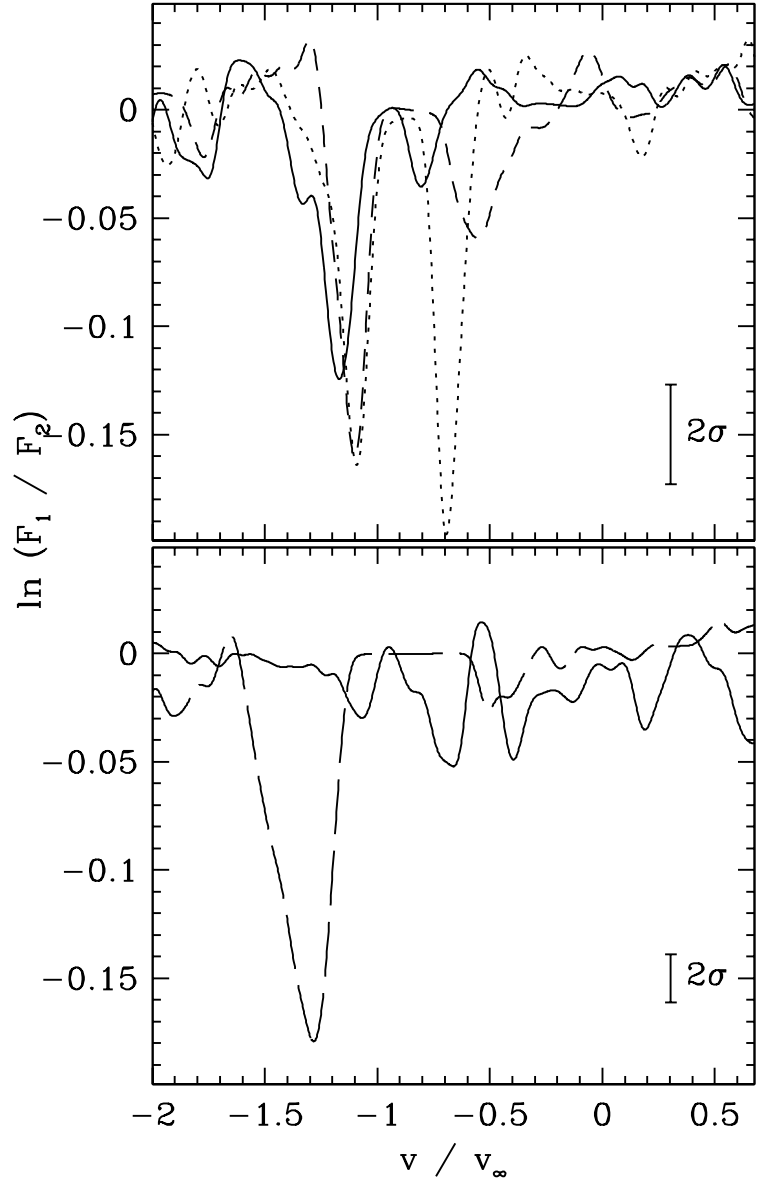}
\caption{Top panel:  the differences of apparent optical depths of the variable details in the profiles of OVI (full line),
S IV at 1062.7 (dotted line)
and  S IV at 1073.0 (dashed line).  Bottom panel: Si III 1108.4 (full line) and C III 1175.7 (long-dashed
line). \label{fig3}}
\end{figure}

\end{document}